\documentclass[twocolumn]{aastex631}

\usepackage{xspace}
\usepackage{xcolor}

\newcommand\Msun{M$_{\odot}$\xspace}

\newcommand\kms{km s$^{-1}$\xspace}

\newcommand{\sn}{SN 2026ahvu}

\shorttitle{SN Ia in NGC 1433}
\shortauthors{Sarbadhicary et al.}

\begin{document}

\title{SN 2026ahvu in NGC 1433 May Be a Prompt Type Ia Supernova with a Delay Time Under 100 Myr}

\author[0000-0002-4781-7291]{Sumit K. Sarbadhicary}
\affiliation{Department of Physics and Astronomy, The Johns Hopkins University, Baltimore, MD 21218 USA}
\author[0000-0001-7448-1749]{Kiana Henny}
\affiliation{Department of Physics and Astronomy, University of Wyoming, Laramie, WY 82071, USA}
\author[0000-0002-2545-1700]{Adam Leroy}
\affiliation{Department of Astronomy, Ohio State University, 180 W. 18th Ave, Columbus, Ohio 43210} \affiliation{Center for Cosmology and Astroparticle Physics, 191 West Woodruff Avenue, Columbus, OH 43210, USA }
\author[0000-0002-4378-8534]{Karin Sandstrom}
\affiliation{Department of Astronomy \& Astrophysics, University of California, San Diego, La Jolla, CA 92093 USA}
\author[0000-0003-3917-6460]{Kirsten~L.~Larson}
\affiliation{AURA for the European Space Agency (ESA), Space Telescope Science Institute, 3700 San Martin Drive, Baltimore, MD 21218, USA}
\author[0000-0002-8528-7340]{David Thilker}
\affiliation{Department of Physics and Astronomy, The Johns Hopkins University, Baltimore, MD 21218, USA}
\author[0000-0002-5259-2314]{Gagandeep~S.~Anand}
\affiliation{Space Telescope Science Institute, 3700 San Martin Drive, Baltimore, MD 21218, USA}
\author[0000-0003-0410-4504]{Ashley~T.~Barnes}\affiliation{European Southern Observatory (ESO), Karl-Schwarzschild-Stra{\ss}e 2, 85748 Garching, Germany}
\author[0000-0002-1790-3148]{Laura A. Lopez}
\affiliation{Department of Astronomy, Ohio State University, 180 W. 18th Ave, Columbus, Ohio 43210} 
\affiliation{Center for Cosmology and Astroparticle Physics, 191 West Woodruff Avenue, Columbus, OH 43210, USA }
\author[0000-0001-6038-9511]{Daniel Maschmann}
\affiliation{Université Côte d'Azur, Observatoire de la Côte d'Azur, CNRS, Laboratoire Lagrange, France}
\author[0009-0002-2740-2993]{Marco Troncoso}
\affiliation{Max Planck Institute for Astronomy, K\"onigstuhl 17, 69117 Heidelberg, Germany}
\author[0000-0003-4770-688X]{Hwihyun Kim}
\affiliation{Gemini Observatory/NSF NOIRLab, 950 N Cherry Ave., Tucson, AZ 85719, USA}
\author[0000-0002-2278-9407]{Janice C. Lee}
\affiliation{Space Telescope Science Institute, 3700 San Martin Drive, Baltimore, MD 21218, USA}
\author[0000-0003-0166-9745]{Frank~Bigiel}
\affiliation{Argelander-Institut f\"ur Astronomie, Universit\"at Bonn, Auf dem H\"ugel 71, 53121 Bonn, Germany}
\affiliation{Cluster of Excellence DYNAVERSE}
\author[0000-0003-0946-6176]{Médéric Boquien}
\affiliation{Université Côte d'Azur, Observatoire de la Côte d'Azur, CNRS, Laboratoire Lagrange, France}
\author[0000-0003-0085-4623]{Rupali~Chandar}
\affiliation{Ritter Astrophysical Research Center, University of Toledo, Toledo, OH, 43606}
\author[0000-0001-8241-7704]{Ryan Chown}
\affiliation{Department of Computer Science and Mathematics, Algoma University, 1520 Queen St E, Sault Ste. Marie, ON P6A 2G4, Canada}
\author[0000-0002-5782-9093]{Daniel~A.~Dale}
\affiliation{Department of Physics and Astronomy, University of Wyoming, Laramie, WY 82071, USA}
\author[0000-0003-1943-723X]{Sinan Deger}
\affiliation{Institute of Astronomy and Kavli Institute for
Cosmology, University of Cambridge, Madingley Road, Cambridge, CB3 0HA, UK}
\author[0000-0002-4755-118X]{Oleg~V.~Egorov}
\affiliation{Astronomisches Rechen-Institut, Zentrum f\"{u}r Astronomie der Universit\"{a}t Heidelberg, M\"{o}nchhofstra\ss e 12-14, D-69120 Heidelberg, Germany}
\author[0000-0001-6708-1317]{Simon C.~O.\ Glover}
\affiliation{Institut f\"ur Theoretische Astrophysik, Zentrum f\"{u}r Astronomie der Universit\"{a}t Heidelberg, Albert-Ueberle-Strasse 2, D-69120 Heidelberg, Germany}
\author[0000-0001-9852-9954]{Oleg Y. Gnedin}
\affiliation{Department of Astronomy, University of Michigan, Ann Arbor, MI 48109, USA}
\author[0000-0002-0560-3172]{Ralf S.\ Klessen}
\affiliation{Universit\"{a}t Heidelberg, Zentrum f\"{u}r Astronomie, Institut f\"{u}r Theoretische Astrophysik, Albert-Ueberle-Str.\ 2, 69120 Heidelberg, Germany}
\affiliation{Universit\"{a}t Heidelberg, Interdisziplin\"{a}res Zentrum f\"{u}r Wissenschaftliches Rechnen, Im Neuenheimer Feld 225, 69120 Heidelberg, Germany}
\author[0000-0002-6056-3425]{Jay Gonz\'alez Lobos}
\affiliation{Max Planck Institute for Astronomy, K\"onigstuhl 17, 69117 Heidelberg, Germany}
\author[0000-0001-9020-1858]{Hao He}
\affiliation{Argelander-Institut für Astronomie, University of Bonn, Auf dem Hügel 71, 53121 Bonn, Germany}
\affiliation{Cluster of Excellence DYNAVERSE}
\author[0000-0002-9723-1088]{Rebecca J. Houghton}
\affiliation{Max Planck Institute for Astronomy, K\"onigstuhl 17, 69117 Heidelberg, Germany}
\author[0000-0003-2721-487X]{Debosmita~Pathak}
\altaffiliation{NSF Graduate Research Fellow}
\affiliation{Department of Astronomy, Ohio State University, 180 W. 18th Ave, Columbus, Ohio 43210} \affiliation{Center for Cosmology and Astroparticle Physics, 191 West Woodruff Avenue, Columbus, OH 43210, USA }
\author[0000-0002-0579-6613]{M. Jimena Rodríguez}
\affiliation{Space Telescope Science Institute, 3700 San Martin Drive, Baltimore, MD 21218, USA}
\affiliation{Instituto de Astrofísica de La Plata, CONICET--UNLP, Paseo del Bosque S/N, B1900FWA La Plata, Argentina }
\author[0000-0002-3933-7676]{Eva Schinnerer}
\affiliation{Max Planck Institute for Astronomy, K\"onigstuhl 17, 69117 Heidelberg, Germany}
\author[0000-0002-2471-8442]{Michael Tucker}
\affiliation{Department of Astronomy, Ohio State University, 180 W. 18th Ave, Columbus, Ohio 43210} \affiliation{Center for Cosmology and Astroparticle Physics, 191 West Woodruff Avenue, Columbus, OH 43210, USA }
\author[0009-0005-8923-558X]{Tony D. Weinbeck}
\affiliation{Department of Physics and Astronomy, University of Wyoming, Laramie, WY 82071, USA}
\author[0000-0002-0012-2142]{Thomas G. Williams}
\affiliation{UK ALMA Regional Centre Node, Jodrell Bank Centre for Astrophysics, Department of Physics and Astronomy, The University of Manchester, Oxford Road, Manchester M13 9PL, UK}
\author[0000-0002-3784-7032]{Bradley C. Whitmore}
\affiliation{Space Telescope Science Institute, 3700 San Martin Drive, Baltimore, MD 21218, USA}
\author[0000-0001-7130-2880]{Leonardo \'Ubeda}
\affiliation{Space Telescope Science Institute, 3700 San Martin Drive, Baltimore, MD 21218, USA}

\begin{abstract}
\sn\ is a Type Ia supernova (SN) that occurred in the nearby galaxy NGC~1433. Using archival Hubble and JWST imaging from the PHANGS survey, we show that the SN lies within 0.19$''$ and 0.26$''$ of two UV-bright star clusters. This $0.19/0.26''$ separation corresponds to $8/11$~pc ($17/24$~pc) at the reported distances of $9.1$ and $18.9$~Mpc, respectively. The probability of finding an unrelated Type Ia SN this close to a cataloged star cluster by chance is only $\sim$0.4\%. Thus, \sn\ is likely the first Type Ia SN that can be associated with relatively young star clusters in a nearby galaxy. From spectral energy distribution fitting using HST+JWST (or HST-only) data, we find that the clusters have age estimates of $37_{-5}^{+3}$ ($57_{-9}^{+14}$) Myr and $95_{-13}^{+15}$ ($88_{-12}^{+14}$) Myr. Association with either cluster would mean that \sn\ should be considered a prompt SN Ia and provide one of the most concrete measurements to date of a prompt Ia delay time. The median age of the younger star cluster is close to the minimum possible delay time for SN Ia progenitors. If the SN is indeed associated with the younger cluster, it would imply that Type Ia SNe can occur almost immediately after white dwarf formation. If the SN is associated with the older cluster, this discovery still presents the first example of a prompt Type Ia SN with delay time under 100~Myr.
\end{abstract}


\section{Introduction}\label{sec:intro}

Type Ia supernovae (SNe) are thermonuclear explosions of carbon-oxygen white dwarfs in binary systems. The nature of the explosion mechanism and binary companion have long been the subject of debate \citep[see][for reviews]{Maoz2014, Ruiter2025}, with proposed channels including single-degenerate (SD) explosions driven by mass accretion onto a white dwarf \citep[e.g.,][]{whelan73, nomoto82} or double-degenerate (DD) explosions driven by the merger of two white dwarfs \citep{iben84, tutukov79, Shen2018}, although evidence has accumulated that SD channels operate in a minority of SNe Ia \citep[e.g.,][]{Schaefer2012, Margutti2012, Silverman2013, Chomiuk2016, Tucker2020}. A consensus on the SNe Ia progenitor channels remains a highly sought-after goal, given their use as standard candles in cosmology \citep[e.g.,][]{phillips1993, Riess1998} and their importance as sources of Fe-group elements in the universe \citep[e.g.,][]{raiteri1996, matteucci2001}.

Two populations of SNe Ia have long been proposed \citep[e.g.,][]{Scannapieco2005, Mannucci2006} -- a prompt (few 100~Myr) and delayed (several Gyr) component -- based on the observed correlation between SN rates and host galaxy colors and star-formation rates \citep[e.g.,][]{Sullivan2006,Li2011,Smith2012}. The SN Ia delay-time distribution (DTD) from extragalactic surveys further supported this bimodality, showing SN rates per M$_{\odot}$ following a consistent $t^{-1}$ power-law distribution across a broad range of delay-times, from 40~Myr to $\gtrsim$10~Gyr  \citep[e.g.][]{Maoz2010, Maoz2011, Graur2011, Maoz2012, Graur2014, Friedmann2018, Strolger2020, Freundlich2021, Wiseman2021}. The slope and the Gyr-long delay-times are naturally reproduced by DD progenitors, whose evolutionary timescales are set by angular momentum loss from gravitational wave emission \citep[e.g.,][]{Totani2008, Toonen2012}. The progenitor composition of the prompt population is less well-understood. Population synthesis models typically predict $\sim$100~Myr as the shortest delay-times for DD channels and SD channels with non-degenerate companions, and about $\sim$40~Myr for SD channels with H-stripped companions \citep[see reviews in][]{Maoz2014, Liu2023}. These predictions can be highly uncertain given various poorly-understood aspects of binary evolution, such as common-envelope evolution \citep{Ruiter2011, Toonen2012, Zapartas2017}. 

Characterization of prompt SNe Ia from observations has been limited. The presence of prompt Ia is only known statistically from DTD surveys. Age information in these surveys often come in broad age-bins \citep[$\gg$100 Myr;][]{Tojeiro2009} from star-formation histories averaged over entire galaxies or cosmological volumes, given the uncertainties associated with integrated photometry and spectra \citep[e.g.,][]{conroy2010, Wuyts2011}. Inferring ages from stellar populations around individual SNe can be challenging  because the region of a galaxy where a SN explodes often blends a range of ages with a complex star formation history \citep[e.g.,][]{Badenes2009,Auchettl2019,Takaro2020}. A definitive association of a Type Ia with a simple stellar population (e.g., star cluster) with a characteristic age is thus necessary, but has remained elusive. Only one known Type Ia (SN 2019ein) has been found near a globular cluster \citep{Bregman2024}, supporting the delayed channel. 

Here, we present the observed association of a very recent Type Ia SN, \sn\ \citep[discovered on 2 Nov 2026 by the ATLAS team][]{}, with two UV-bright, young stellar clusters in the galaxy NGC~1433. At a distance of either $9.1 \pm 1.0$ \citep{Sabbi2018} or 18.9$^{+0.4}_{-0.6} $ Mpc \citep{Scheuermann2022} (see below), this galaxy is sufficiently nearby that star clusters can be identified with Hubble and JWST imaging and isolated from the background distribution of stars in the galaxy. Spectral energy distribution modeling of UV through near-IR photometry places strong constraints on the ages and masses of the clusters \citep[e.g.,][]{Thilker2025,Henny2025}, thereby suggesting that \sn\ is likely a prompt SN Ia.  

\section{Data}\label{sec:data}

\subsection{SN Detection \& Classification}

\sn\ (ATLAS26ncp) was discovered\footnote{\url{https://www.wis-tns.org/astronotes/astronote/2026-310}} 2 Nov 2026 by the ATLAS SN search \citep{Srivastav2026}. It is located at coordinates RA 3$^{\rm h}$ 41$^{\rm m}$ 55$\fs$511, Dec $-$47$^{\circ}$ 13$'$ 39$\farcs$39, which falls within the disk of the nearby galaxy NGC~1433. The SN was classified as a Type Ia based on follow up spectroscopy with Keck/DEIMOS\footnote{\url{https://www.wis-tns.org/astronotes/astronote/2026-311}}.

Per \citet{Tonry2018a}, ATLAS has positional accuracy of $\approx 0.07''$ for $17$~mag stars and is tied to the \textit{Gaia} reference frame \citep{Tonry2018b}. Since the discovery magnitude of ATLAS26ncp is $\approx 17$~mag, a position accuracy of $< 0.1''$ is reasonable but note that there is not yet a formal reported uncertainty on the position. The HST and JWST data that we compare to are also referenced to the Gaia frame, so we do not expect any systematic offset.

\begin{figure*}
    \centering
    \includegraphics[width=\textwidth]{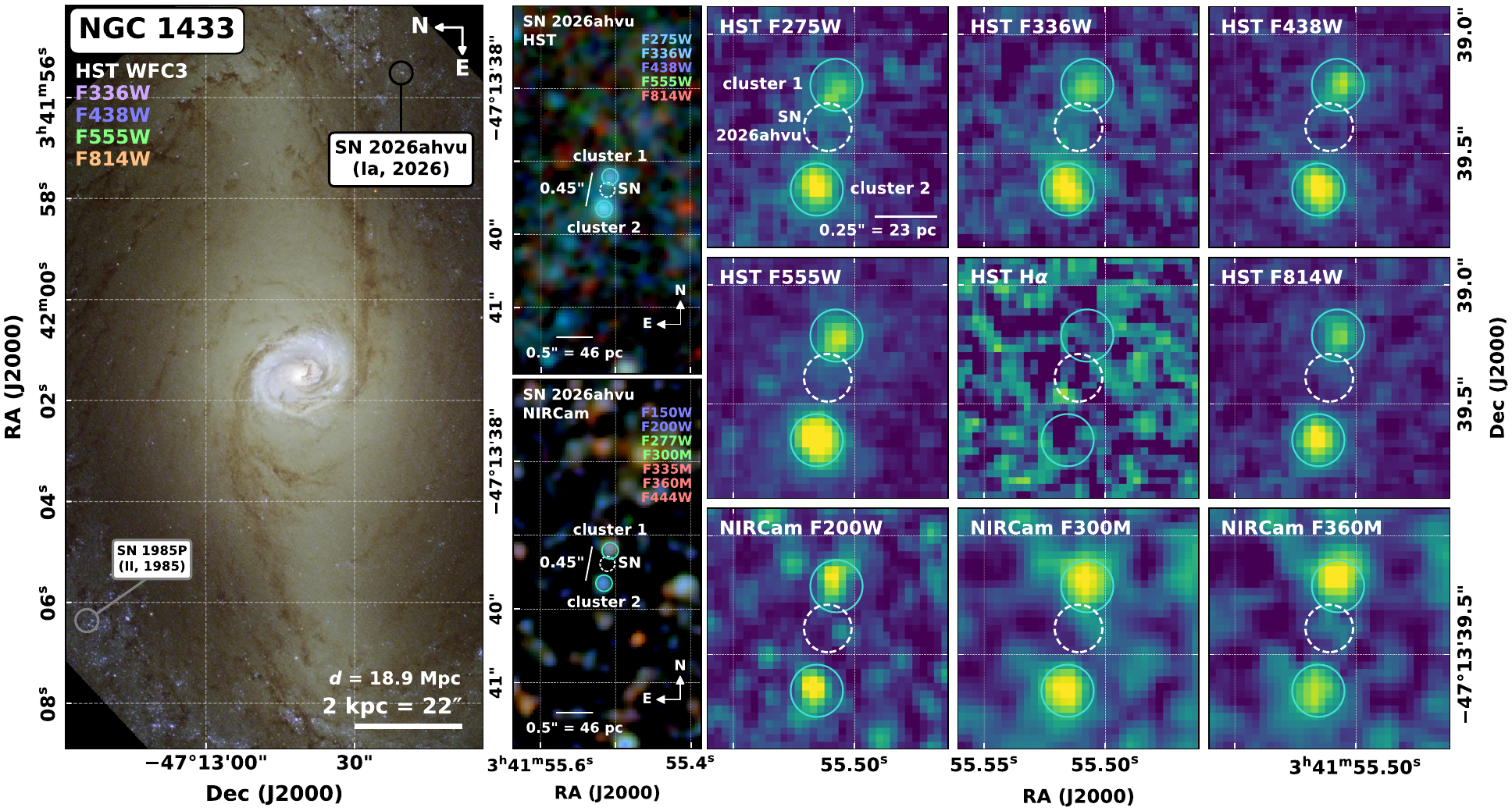}
    \caption{\textbf{Location of \sn\ in NGC\,1433.} \textit{Left:} HST/WFC3 color image of the bar and nuclear ring of NGC\,1433 (credit: ESA/Hubble \& NASA; F336W, F438W, F555W and F814W), rotated by 90$^\circ$. The black circle marks the TNS position of \sn\ (Type Ia). The grey circle marks SN\,1985P (Type II), another historical SN in this galaxy within our 5$'$ footprint. The scale bar is $\approx 1$~kpc at $9.1$~Mpc or $\approx 2$\,kpc ($22''$) at $d = 18.9$\,Mpc. \textit{Middle:} $2.5'' \times 5''$ north-up color cutouts centered on \sn, from HST/UVIS (top; F275W+F336W, F438W, F555W, F814W) and JWST/NIRCam (bottom; F150W+F200W, F277W+F300M, F335M+F360M+F444W). The dashed white circle shows the $\sim$1$\sigma$ ($0.1''$) positional uncertainty of \sn. The cyan circles mark the two candidate host star clusters, $0.19''$ north (cluster~1) and $0.26''$ south (cluster~2) of the SN. The clusters are $0.45''$ apart, $\approx 20$~pc at 9.1 Mpc or 41~pc at 18.9 Mpc. \textit{Right:} $1'' \times 1''$ single-band cutouts on a common north-up grid, in order of wavelength: HST UV--optical, continuum-subtracted HST H$\alpha$, and JWST/NIRCam. Each panel uses an asinh stretch from $-1\sigma$ to its peak, clipped to 5--30$\sigma$. The markers are the same as in the middle panels, and the scale bar is $0.25''$ (11 or 23\,pc, respectively). Both clusters are detected from the UV to $\sim$4.4\,$\mu$m. Neither cluster shows H$\alpha$ emission, and neither is detected by MIRI (not shown): the clusters lie in a gap between dust/PAH filaments.}
    \label{fig:fig1}
\end{figure*}

\subsection{Archival HST \& JWST data}
NGC 1433 was observed with VLT-MUSE and JWST as part of the PHANGS survey \citep{Emsellem2022,Lee2023}.  Observations with HST were originally obtained by the LEGUS survey \citep{calzetti2015}, and re-processed through the PHANGS-HST star cluster pipeline \citep{Lee2023}. Observations with photometric filters span the near-UV to near-IR, including HST F275W, F336W, F438W, F555W, F814W (plus narrow-band F657N, used for H$\alpha$), and JWST F200W, F300M, F335M, F360M. The JWST observations and processing are described in \citet{Lee2023} and \citet{williams2024}. 
We also use the narrow band H$\alpha$ map of NGC~1433 from \citet{Chandar2025}, constructed from WFC3/UVIS F657N narrow-band imaging. In the figure we show F657N with the stellar continuum estimated from the broad-band HST filters and subtracted and we model the F657N flux with no continuum subtraction. The continuum-subtracted HST H$\alpha$ fluxes are corrected for [N\,\textsc{ii}] contamination and dust extinction using the Balmer decrement computed using emission-line maps from VLT/MUSE \citep{Emsellem2022}. The Hubble observations were used to construct catalogs of star clusters \citep{Maschmann2024, Thilker2025}, and then their properties are estimated based on SED modeling with CIGALE \citep{boquien2019} and association with H$\alpha$ emission \citep{Thilker2025}. The images and cluster catalogs are publicly available\footnote{\url{https://archive.stsci.edu/hlsp/phangs}}.

\subsection{Distance to NGC 1433}

\citet{Sabbi2018} reported a distance to NGC 1433 of 9.1$\pm$1.0 Mpc via the tip of the red giant branch (TRGB) using observations from HST as part of the LEGUS program. However, they note in their text that the distance estimates for five galaxies in their sample, including NGC 1433, are ``quite uncertain'' and should be used with caution. \citet{Scheuermann2022} used the planetary nebula luminosity function (PNLF) to measure a distance of 18.94$^{+0.4}_{-0.6}$~Mpc. The imminent peak of the SN Ia light curve will provide a swift resolution to any distance controversy. We proceed quoting results for both distances wherever relevant. The distance affects the estimated stellar mass of our clusters, which is not key to any of our conclusions, and the projected physical separation between \sn\ and the nearby clusters. For 9.1 Mpc, $1'' = 44$~pc, and for $18.9$ Mpc, $1'' = 92$~pc.

\section{Results}

\subsection{Association of \sn\ with the Clusters}

Figure~\ref{fig:fig1} shows a compilation of photometric data with the position of \sn\ overlaid. Two nearby star clusters are evident to the north (Cluster 1) and south (Cluster 2) of the SN position. The SN appears to be slightly closer to Cluster 1 at $0.19''$ than Cluster 2 at $0.26''$, but the difference is comparable to the nominal $\approx 0.1''$ astrometric uncertainty in the ATLAS images. At the 9.1 (18.9)~Mpc distance of NGC~1433, these separations correspond to $0.19'' \approx 8$ ($18$) pc and $0.26'' \approx 11$ ($24$) pc, respectively.

Cluster 1 is known and included in the PHANGS-HST DR5 cluster catalog, having been identified both by visual inspection and by machine learning classification \citep{Maschmann2024}. Its physical properties have been modeled by \citet{Thilker2025} and \citet{Henny2025}, which we revisit below. Cluster 2 was missed in the DR5 catalog, though its properties when measured manually indicate that it should have been included. 

We confirm both objects have morphologies consistent with star clusters of Class 2 (asymmetric, single peak, centrally concentrated). They have multiple concentration indices MCI(in)$\sim$0.1 and MCI(out)$\sim$0, values which are associated with extended clusters rather than individual stars in the PHANGS galaxies \citep[see][for definitions of MCI and discussion its use to identify clusters]{Thilker2022}. In tests that accompanied the training of the machine learning algorithm used to produce the cluster catalogs, the algorithm reproduced human classification $\approx 70-80\%$ of the time for Class 1 (symmetric) clusters and $\approx 40-70\%$ for Class 2 (asymmetric) clusters \citep{Whitmore2021}. Completeness tests also showed that the catalogs are $\approx 70\%$ complete
down to $m_{\rm V}\approx24.0$ \citep{Whitmore2021}. 

We estimate the probability of the SN being associated with the older stellar population and lying near the clusters by chance. We treat the JWST F300M ($\lambda = 3~\mu$m) image, which traces the stellar mass distribution, as a proxy for the probability of SNe Ia associated with the older stellar population exploding at a given location. Within the overlap of the HST and JWST footprints, only $\approx 0.4\%$ of the F300M flux lies within $0.25''$ of a Class 1+2 ML cluster in the \citet{Maschmann2024} catalog (i.e., within a distance closer than what we observe). Similarly, only $\approx 0.3\%$ of the total area in the overlap of the surveys lies within $0.25''$ of a cluster.

The fact that the clusters are UV-bright drives our conclusion that \sn\ is likely a prompt SN Ia. If we consider only proximity to any bright (S/N$>5$) near-UV (F275W) emission, rather than distance to identified clusters, the probability of chance alignment is somewhat higher. After excluding the starburst center, we estimate that $\approx 4.6\%$ of the area and $\approx 3.7\%$ of the F300M flux within the JWST+HST survey area lie within $0.25''$ of a pixel with significant NUV emission (this high S/N near-UV emission follows the structure of NGC 1433 and so appears to not be significantly contaminated by Milky Way foreground stars). This represents a conservative limit on the chance association of a SN with any pixel with NUV emission, however the association with known clusters is far less likely.

\subsection{Cluster Age \& Mass from SED fitting}\label{sec:sedfit}
\begin{figure*}
    \centering
    \includegraphics[width=\textwidth]{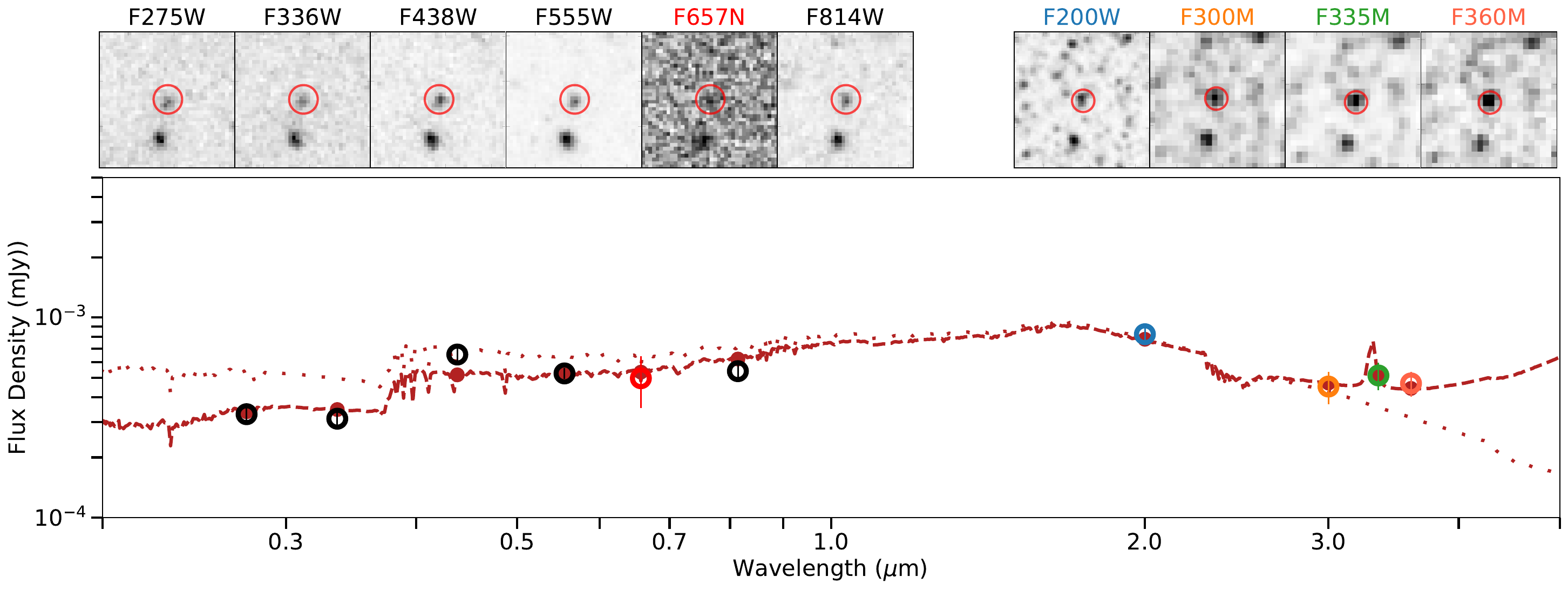}
    \includegraphics[width=\textwidth]{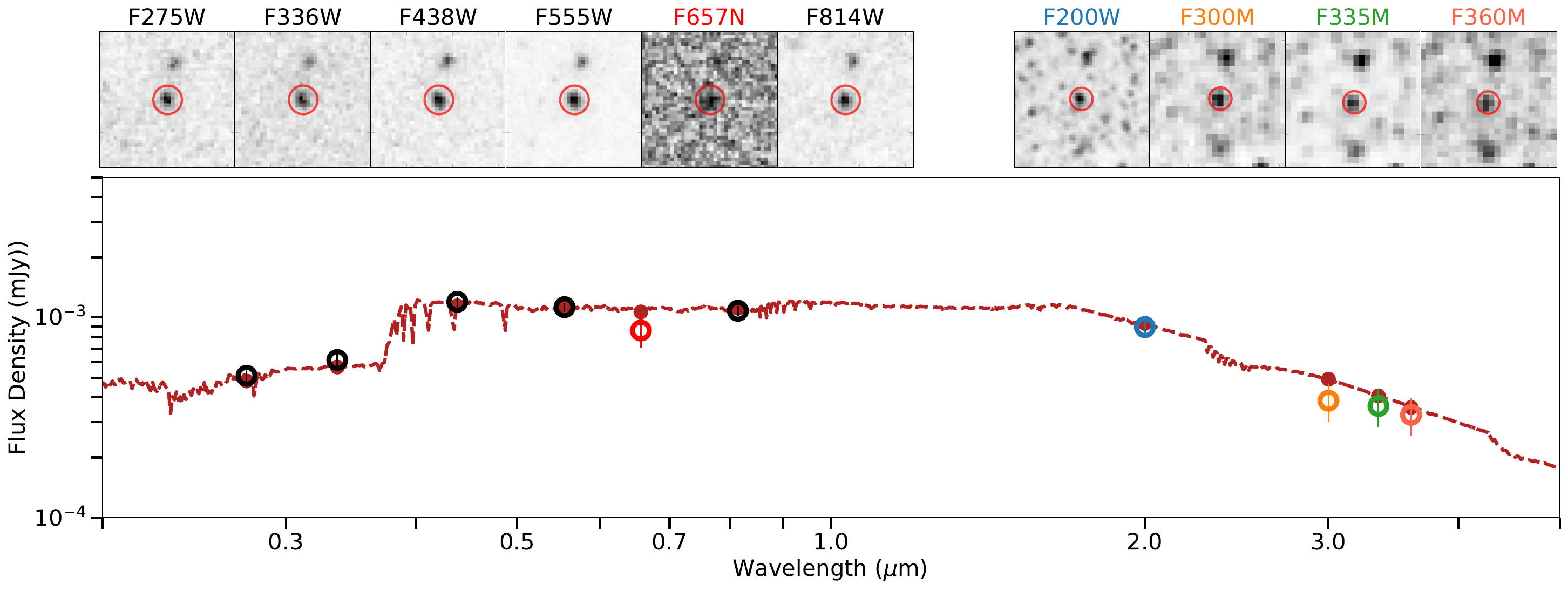}
    \caption{Observed SED (open circles with error bars, see Table \ref{tab:photandfit}) with their best-fit modeled spectra (dashed dark red line) for Cluster 1 (top) and Cluster 2 (bottom). Solid dark red points show the modeled fluxes in each filter. The dotted line in Cluster 1 indicates the modeled intrinsic stellar continuum, without dust absorption or emission. Note the divergence of Cluster 1's NIR modeled spectrum from the pure stellar continuum, indicating that there must be some hot dust present. Best fit ages and masses are reported in Table \ref{tab:photandfit}, along with 16-84\% confidence ranges. The presence of bright near-UV emission in both clusters leads to fitted ages $\lesssim 100$~Myr and suggests that \sn\ is a prompt SN Ia if it is associated with either cluster.}
    \label{fig:clustersed}
\end{figure*}

In this section, we measure the possible ages and masses of the two clusters near the SN using the full HST and NIRCam bands available in PHANGS. We quote measurements using both HST+NIRCam, as well as HST-only (excluding NIRCam) to account for stellar population modeling uncertainties in the near-IR \citep[e.g][]{Henny2025, Pedrini2025}.

We measure NUV-to-NIR photometry with HST and NIRCam imaging including the HST bands F275W, F336W, F438W, F555W, F657N, F814W and the NIRCam bands F200W, F300M, F335M, and F360M. Then we model the SED of each cluster using CIGALE \citep{boquien2019} to estimate ages, reddening and masses. Our approach follows the method described in \cite{Henny2025}, and that paper includes discussion of the merits and complications of NUV-optical-NIR SED fitting and a comparison of this approach with HST-only SED modeling. Note that we incorporate the F657N filter directly in SED modeling without performing continuum subtraction or correction for [NII] emission and dust extinction. Also note that we do not use MIRI filters in our SED modeling because the principle of energy balance is not clearly satisfied on these scales. 

We perform aperture photometry on the two clusters using a fixed aperture with radius $r=$4-times the HST pixel size (i.e., $4 \times 0.03962 = 0.158''$) for all HST bands and the JWST 3, 3.3 and 3.6~$\mu$m bands. We use an aperture with radius equal to 4 NIRCam short pixels (i.e., $4 \times 0.031 = 0.124''$) for the F200W band. This HST+NIRCam long wavelength aperture corresponds to a physical diameter of 29~pc (18.94~Mpc)/14~pc (9.1~Mpc). We estimate the sky value as a robust median, using sigma clipping to reduce the impact of neighbors, perform background subtraction using a fixed 7-9 pixel annulus, and account for foreground reddening. This technique, which is described in \citet[][in press]{maschmann2026} and tested on star clusters in the optical regime in Henny et al. (in prep.), mitigates the effects of nearby source contamination while also allowing us to determine custom aperture corrections. 

We determine aperture corrections by measuring the profiles of the clusters in the F555W band with 12 slices bisecting the cluster at evenly spaced angles. Then we stack these one-dimensional profiles, find the median profile, and fit them with a Gaussian convolved with a Moffat-EFF \citep{eff1987}. The sizes inferred from this approach agree to within 10\% with cluster radii determined via star density in Local Group Galaxies (Henny et al. in prep., \citealt{johnson2022}). We find F555W aperture correction factors of 1.18 and 1.19 respectively. These are lower compared to the correction factor of 1.77 derived from bright, isolated star clusters in NGC 1433 found in \cite{deger2022}, with our measured corrections expected to represent a significant improvement over these previous values.

These two clusters do not have measurable nebular emission, as indicated by their lack of elevated F657N. Nor do they have NUV-optical colors consistent with old globular clusters \citep{Thilker2025}. Therefore, we adopt a modeling setup that includes stellar emission \citep{bruzual2003}, dust absorption \citep{cardelli1989}, and dust emission \citep{draine2014}\footnote{We eliminate models that have $\gamma = 0$ and an age less than 10~Myr. In the \citet{draine2014} models, $\gamma$ captures the fraction of dust mass heated by a power law of ``PDR-like'' radiation fields rather than the diffuse interstellar radiation field. Based on past experience, this avoids the unphysical case where CIGALE tends to prefer nonphysical dust emission scenarios where young modeled clusters also produce no high-energy photons which can heat dust.}, but not nebular emission. We use \cite{bruzual2003} stellar tracks with a metallicity of Z=0.02 (i.e., Z$_{\odot}$), an age grid of 1-10~Myr with $\Delta$ T = 1~Myr and 11-13750~Myr with $\Delta$ log (T/Myr) = 0.3, and an attenuation grid of A$_{550}$ = [0.01:30] with 100 log-spaced values, plus zero. We use a \cite{chabrier2003} IMF. There are limited metallicities available in any SSP model, and the two closest metallicities we have available are Z=0.02, and Z=0.01. The assumption of Solar metallicity is in reasonable agreement with the measured nebular abundance gradient in \citet{Groves2023}, which yields a slightly subsolar value ($12+\log_{10} {\rm O/H} \approx 8.5{-}8.6$) near the location of the SN.

Figure \ref{fig:clustersed} shows the two clusters' observed SEDs and best-fit modeled spectra.  The best fit age, mass, extinction are reported in Table \ref{tab:photandfit}, along with the 16 to 84\% range of each parameter. We find that the clusters 1 and 2 have age estimates of $37_{-5}^{+3}$ ($57_{-9}^{+14}$) Myr and $95_{-13}^{+15}$ ($88_{-12}^{+14}$) Myr, respectively, for the HST+NIRCam (HST-only) fits. Cluster 2 also appears to be roughly 2-4 times more massive than cluster 1. Both clusters appear to exhibit minimal reddening.

As noted above, neither cluster exhibits any nebular emission, as indicated by the lack of H$\alpha$ emission above the stellar continuum in Figure \ref{fig:fig1} and the lack of excess F657N in Figure \ref{fig:clustersed}. They also do not have associated counterparts in MIRI bands (7.7$\mu$m, 10$\mu$m, 11.3$\mu$m, 21$\mu$m), and the MIR only exhibits diffuse dust that is not aligned with the clusters' locations. Cluster 1 exhibits a small amount of dust emission, as indicated by the flat slope between F300M and F360M and slight elevation of F335M, indicating very weak 3.3~$\mu$m PAH emission. Cluster 2 exhibits a red optical slope and steeply declining NIR slope, consistent with no reddening.

Because these clusters have little or no associated dust, their optical slopes between the NUV and B bands can be attributed to the 4000\AA\ break, rather than being an effect of reddening, indicating that they are older than 10~Myr. Similarly, their bright UV and B band fluxes indicate that they are less than a few 100~Myr old. These indications provide broad age bounds even without SED modeling.

Our models do not account for the effects of stochastic sampling of the stellar initial mass function. This might be expected to be relevant given the low mass of the clusters, especially at the near distance \citep[e.g.,][]{dasilva2014}. However, note that the most likely sense of bias is that we overestimate the age of the clusters in the case that they failed to produce massive young, UV-bright stars. We expect that future work will investigate the impact of stochastic sampling.

\begin{deluxetable}{lcc}[t!]
\centering
\tabletypesize{\small}
\tablecaption{Photometry and SED Fitting \label{tab:photandfit}}
\tablewidth{0pt}
\tablehead{
\colhead{Quantity} & 
\colhead{\textbf{Cluster 1}} & 
\colhead{\textbf{Cluster 2}} 
}
\startdata
\multicolumn{3}{c}{Photometry} \\
\multicolumn{3}{c}{($10^{-3}$ mJy)} \\
\hline
\hline
F275W & $0.329 \pm 0.023$ & $0.512 \pm 0.026$ \\
F336W & $0.312 \pm 0.029$ & $0.613 \pm 0.031$ \\
F438W & $0.653 \pm 0.036$ & $1.199 \pm 0.037$ \\
F555W & $0.526 \pm 0.019$ & $1.127 \pm 0.021$ \\
F657N & $0.498 \pm 0.142$ & $0.861 \pm 0.148$ \\
F814W & $0.540 \pm 0.033$ & $1.081 \pm 0.034$ \\
F200W & $1.068 \pm 0.056$ & $1.158 \pm 0.061$ \\
F300M & $0.543 \pm 0.080$ & $0.462 \pm 0.080$ \\
F335M & $0.584 \pm 0.076$ & $0.411 \pm 0.077$ \\
F360M & $0.528 \pm 0.063$ & $0.368 \pm 0.067$ \\
\hline
\hline
\multicolumn{3}{c}{Best fit properties \textbf{(HST+JWST)}} \\
\hline
\hline
Age (Myr) & & \\
... best fit & 34 & 110 \\
... (16, 50, 84) & (32, 37, 40) & (82, 95, 110) \\
$E(B-V)$ (mag) &  & \\
... best fit & 0.07 & 0.0 \\
... (16, 50, 84) & (0.056, 0.069, & (0.003, 0.005, \\
 & 0.084) & 0.007) \\
Mass [10$^3$ M$_\odot$] & & \\
... at 9.1 Mpc & 0.74 & 2.6\\
... at 18.9 Mpc & 3.2 & 11.4 \\
\hline
\hline
\multicolumn{3}{c}{Best fit properties \textbf{(HST only)}} \\
\hline
Age (Myr) & & \\
... best fit & 71 & 95 \\
... (16, 50, 84) & (49, 57, 71) & (76, 88, 102) \\
$E(B-V)$ (mag) & & \\
... best fit & 0.0 & 0.0 \\
... (16, 50, 84) & (0.0043, 0.0091 & (0.0044, 0.0096 \\
 & 0.022) & 0.022) \\
Mass [10$^3$ M$_\odot$] & & \\
... at 9.1 Mpc & 0.95 & 2.2 \\
... at 18.9 Mpc & 4.1 & 9.6 \\
\hline
\enddata
\end{deluxetable}

\section{Discussion}\label{sec:disc}

The presence of two young ($<$100 Myr) clusters and the low probability of a chance coincidence indicate that this is a SN Ia from the putative ``prompt'' channel of SNe Ia. The presence of a significant SN Ia population with these delay-times has been indicated in statistical DTD measurements of extragalactic SN Ia samples \citep[e.g.,][]{Maoz2010, Maoz2014}, SN remnants (SNRs) in the Local Group \citep[e.g.,][]{MaozBadenes2010}, and chemical evolution studies \citep[e.g.,][]{Dubay2024}. Local star-formation histories from resolved stellar populations around SNRs have also hinted at the presence of a prompt channel for SNe Ia \citep{Badenes2009, MaozBadenes2010}; in particular, the LMC SNR N103B, which is within a projected 40~pc of the cluster NGC~1850 \citep{Li2017} with a prominent peak of star formation 50$-$100~Myr ago \citep{Badenes2009,Nied15}. \textit{To our knowledge, this work is the first definitive discovery of a prompt SN Ia based on age dating of its associated star cluster} as previous nearby SNe Ia have not been associated with young clusters or progenitor systems (e.g., for SN~2011fe in M101: \citealt{Li2011fe}).

This is also one of the first associations of a prompt SN Ia with an individual stellar cluster. To our knowledge the only previous candidate cluster{-}SN Ia association was SN 2019ein in NGC~5353, where the SN was located within $\sim$59~pc of a candidate globular cluster \citep{Bregman2024}. While an age of that cluster was not reported, typically such globular clusters are older than several Gyr, making 2019ein more consistent with the delayed SN Ia channel. 

In theory, prompt SNe Ia can be produced in white dwarf binaries with both non-degenerate (e.g., He-envelope after H is stripped) or degenerate (another white dwarf) companions \citep{Ruiter2025}. Delay times in the former channel are mainly dictated by the companion lifetime \citep[30--100~Myr;][]{wang2009,wang2014}, while the latter is dictated by the gravitational in-spiral time. White dwarfs are typically produced by stars $<8$ \Msun, with lifetimes of 40 Myr (at Z$_{\odot}$), which sets an effective lower bound on the SN Ia delay-time.

If \sn\ is associated with Cluster 1, then the young age of $37$~Myr (posterior confidence interval \mbox{32--40~Myr})
implies an almost immediate supernova after formation of the WD. This scenario comes in tension with the low cluster mass (Table~\ref{tab:photandfit}), which may limit the formation of massive stars due to stochastic sampling of the IMF, and the projected distance of \sn\ at 5 times the half-light radius of the cluster (0.036\arcsec\ in F555W). Cluster 2 with an age of $95$~Myr (confidence interval 82--110~Myr) is older and more massive, and therefore, a more natural birthplace. 

The massive binary progenitor of the SN likely formed near the core of the cluster, where it could experience gravitational interactions with another binary or a stellar-mass black hole. Strong encounters with a massive perturber typically result in velocity kicks sufficient to escape the cluster \citep[e.g.,][]{cabrera2023, weatherford2023}. Given the estimated cluster mass $\sim$10$^4$~\Msun and half-mass radius $\sim$4~pc, the escape speed from the core is $\sim$5~\kms. At such speed, the progenitor binary could reach its present projected location within a few Myr. Such a velocity kick is smaller than an expected orbital speed of the progenitor binary ($\sim$100~\kms), which must be very compact to produce the SN and therefore would survive the encounter.

The association of \sn\ with either of these clusters provides strong evidence that prompt Ia progenitor binaries indeed exist in nature. As the SN gets brighter, and photometric and spectroscopic observations accumulate, more light will be shed on the nature of progenitor system, and it is important to interpret the findings in the context of the host cluster system.

\section{Conclusion}

We present the association of \sn, a recently discovered Type Ia SN, with two young star clusters in the nearby galaxy NGC~1433. The probability of a chance alignment of a SN associated with the old stellar population with a cataloged cluster would be small, $\sim0.4$\%. The probability of a SN associated with the old stellar population being so close to any detected near-UV emission would be $\sim$4\%. The two clusters are both notably blue, with bright emission particularly in the near-UV filters covered by {\em Hubble}, however they have no counterparts in dust emission or in H$\alpha$ emission, indicating that they are $>10$~Myr old. 

We perform SED fitting on the near-UV to near-IR photometry of both clusters using techniques described in multiple papers by the PHANGS team \citep{Thilker2022, Maschmann2024, Thilker2025, Henny2025}. Our best measurements and photometry are summarized in Table \ref{tab:photandfit}. Fits with the HST+JWST data (or HST-only data) indicate the northern cluster, which is closest to the SN, has an age of $37_{-5}^{+3}$ ($57_{-9}^{+14}$) Myr, while the southern cluster has an age of $95_{-13}^{+15}$ ( $88_{-12}^{+14}$) Myr. These ages do not depend on the uncertain distance to NGC 1433. The best fit Cluster 1 (Cluster 2) masses based on HST+NIRCam SED analysis are 0.74 (2.6) $\times 10^3$~M$_\odot$ for a distance of $9.1$~Mpc and 3.2 (11.4) $\times 10^3$~M$_\odot$ for $18.9$~Mpc.

Given the association with the cluster by proximity, and cluster ages in the 40$\sim$100 Myr range, \sn\ may be the first confirmed SN from the prompt Ia channel. More observations will accumulate in the coming weeks as the SN brightens, providing independent constraints from multi-wavelength light curves and spectroscopy and refining the best fit position and uncertainty in the position. We encourage analyzing these in the context of the SN host cluster information presented here, and exploring the parameter space of the various SN Ia progenitor routes from population synthesis models that fit these ages.

\section*{Acknowledgments} \label{sec:acknowledgments}

This work has been carried out as part of the PHANGS collaboration, based on observations made with the NASA/ESA/CSA Hubble and James Webb Space Telescopes. The data were obtained from the Mikulski Archive for Space Telescopes at the Space Telescope Science Institute, which is operated by the Association of Universities for Research in Astronomy, Inc., under NASA contract NAS 5-26555 for HST and NAS 5-03127 for JWST. These observations are associated with JWST GO program 2107 and HST GO programs 13364, 15654, 17126, and 17457. 

AKL gratefully acknowledges support from JWST-GO-03707.001-A and JWST-GO-04793.002-A. KS acknowledges support from JWST-GO-03707.005-A and JWST-GO-04793.005-A. LAL gratefully acknowledge financial support from the Heising-Simons Foundation 2026-6821. OYG gratefully acknowledges support from JWST-AR-08735.

MB acknowledges support by the ANID BASAL project FB210003. This work was supported by the French government through the France 2030 investment plan managed by the National Research Agency (ANR), as part of the Initiative of Excellence of Université Côte d’Azur under reference No. ANR-15-IDEX-01. This research was funded, in whole or in part, by the French National Research Agency (ANR), grant ANR-24-CE92-0044 (project STARCLUSTERS).

OE acknowledges funding from the Deutsche Forschungsgemeinschaft (DFG, German Research Foundation) -- project-ID 541068876.

JGL and RJH acknowledge support from the DLR (German Aerospace Agency) via grant 50 OR2401. 

FB and HH acknowledge funding by the Deutsche Forschungsgemeinschaft (DFG, German Research Foundation) under Germany’s Excellence Strategy EXC Our Dynamic Universe and gratefully acknowledges the Collaborative Research Center (SFB 1601 sub-project B3) funded by the Deutsche Forschungsgemeinschaft (DFG, German Research Foundation)

\facilities{JWST, HST}

\software{astropy \citep{ASTROPY13,ASTROPY18}}

\bibliography{main}{}
\bibliographystyle{aasjournal}


\end{document}